\documentclass[pdftex,twocolumn,epjc3]{svjour3}          

\RequirePackage[T1]{fontenc}
\usepackage{amsmath,mathtools}
\usepackage{lipsum,widetext}
\usepackage{xcolor}

\smartqed  

\RequirePackage{graphicx}
\RequirePackage{mathptmx}      
\RequirePackage{flushend}
\RequirePackage[numbers,sort&compress]{natbib}
\RequirePackage[colorlinks,citecolor=red,urlcolor=blue,linkcolor=blue]{hyperref}
\journalname{Eur. Phys. J. C}

\begin{document}

\title{OSIRIS: A New Code for Ray Tracing Around Compact Objects\thanksref{t1}}


\author{
        Velásquez-Cadavid  J.M.\thanksref{e1,addr1}
        \and
        Arrieta-Villamizar J.A. \thanksref{e2,addr1}
        \and
        Lora-Clavijo, F.D. \thanksref{e3,addr1}
        \and
        Pimentel, O.M. \thanksref{e4,addr1}
        \and
        Osorio-Vargas J.E. \thanksref{e5,addr1}
}

\thankstext[$\star$]{t1}{Thanks to the title}
\thankstext{e1}{e-mail:juan2208056@correo.uis.edu.co}
\thankstext{e2}{e-mail:jesus2208058@correo.uis.edu.co}
\thankstext{e3}{e-mail:fadulora@uis.edu.co}
\thankstext{e5}{e-mail:josorvar@correo.uis.edu.co}
\thankstext{e4}{e-mail:oscar.pimentel@correo.uis.edu.co}

\institute{Escuela de Física, Universidad Industrial de Santander, A. A. 678, Bucaramanga 680002, Colombia \label{addr1}
}

\date{Received: date / Accepted: date}

\maketitle
\begin{abstract}
The radiation observed in quasars and active galactic nuclei is mainly produced by a relativistic plasma orbiting close to the black hole event horizon, where strong gravitational effects are relevant. The observational data of such systems can be compared with theoretical models to infer the black hole and plasma properties. In the comparison process, ray-tracing algorithms are essential to computing the trajectories followed by the photons from the source to our telescopes. In this paper, we present \texttt{OSIRIS}: a new stable FORTRAN code capable of efficiently computing null-geodesics around compact objects, including general relativistic effects such as gravitational lensing, redshift, and relativistic boosting. The algorithm is based on the Hamiltonian formulation and uses different integration schemes to evolve null-geodesics while tracking the error in the Hamiltonian constrain to ensure physical results. We found from an error analysis that the integration schemes are all stable, and the best one maintains an error below $10^{-11}$. Particularly, to test the robustness and ability of the code to evolve geodesics in curved space-time,  we compute the shadow and Einstein rings of a Kerr black hole with different rotation parameters and obtain the image of a thin Keplerian accretion disk around a Schwarzschild black hole. Although \texttt{OSIRIS} is parallelized neither with MPI nor with CUDA, the computation times are of the same order as those reported by other codes with these types of parallel computing platforms.   
\end{abstract}

\section{Introduction}

The study of the null geodesics in the curved space-time generated by a black hole or any other compact object serves to analyze the strong gravity effects that are expected near to them as well as to understand the dynamics of the accretion processes. In 2019, the Event Horizon Telescope collaboration published the first images-ever of the shadow produced by the supermassive black hole at the center of the M87 galaxy \cite{collaboration2019first}. M87$^{*}$ is one of the largest known supermassive black holes and is located at the center of the gargantuan elliptical galaxy, 53 million light-years away. The estimated mass for the black hole is $M_{BH} = (6.5 \pm 0.7) \times 10^{9} M_{\odot}$ and was obtained by comparing the images with an extensive library of ray-traced general-relativistic MHD simulations. 

Through numerical simulations, it is possible to process, analyze and compare the observational data. That is one of the reasons why a large number of ray-tracing codes in general relativistic space-times have been developed. From these codes, it has been possible to simulate the motion of photons around compact objects and thus get a better understanding of the gravitational field exerted over them.
Solving the equations of the null geodesics analytically allows studying spherical orbits of photons that constitute the well-known photon ring \cite{bardeen, teo, liu_2019, gralla}, which conform to the apparent edge of the event horizon and delimit the black hole shadow. 
However, obtaining an analytical solution for an arbitrary space time is not always possible. Therefore, it is necessary to construct numerical codes that allow finding solutions to the geodesic equations for a large number of photons. Based on this, several numerical studies, related to ray-tracing, have been carried out \cite{fanton_1977, luminet, mark_1996, Fuerst_2004, muller_2012, Bohn_2015, younsi_2016, Vincent_2016}, distinguishing in the literature codes such as GeoKerr \cite{dexter}, Gyoto \cite{vincent2011gyoto}, Ray \cite{Dimitrios}, GRay \cite{chan2013gray}, GeoVis \cite{muller_2014}, Pyhole \cite{cunha_2016},     Oddisey \cite{pu2016odyssey},  among others. Each of these stands out for its different attractions, be it for the speed of its algorithms, the ability to solve the geodesic equation in general space-times (stationary or dynamic) in different formalisms, having a user-friendly interface design, or the powerful architecture in which they are written.

Even though nowadays the codes are extremely advanced, including polarized radiative transfer in general relativity \cite{gammie2012formalism, dexter2016public,moscibrodzka2018ipole,pihajoki2018general}, in in this work we present and certify our code \texttt{OSIRIS} ({\bf O}rbits and {\bf S}hadows {\bf I}n {\bf R}elativ{\bf I}stic {\bf S}pace-times), which in its first version focuses on the solution of null geodesics equations, using the Hamiltonian formulation. \texttt{OSIRIS} uses the Backward ray-tracing method to efficiently calculate the shadow, the Einstein ring of a black hole, and different general relativistic effects, such as gravitational lensing, redshift, and relativistic boosting.    Particularly, \texttt{OSIRIS} has been used to study the shadow produced by a naked singularity described by the q-metric, which is the simplest static and axially symmetric solution of Einstein equations with a non-vanishing quadrupole moment \cite{arrieta2020shadows}. The validation of the code has been tested through different applications like the Black hole shadow in Kerr space-time and the shadow produced by a thin accretion disk.

This article is organized as follows. In Section \ref{sec:meq}, we briefly describe the Hamiltonian formulation and the motion equations used to compute the null geodesics. Additionally, we show the backward ray-tracing method to visualize the effects of the strong gravitational field around a compact object. Moreover, to obtain the shadow produced by a black hole, we present the impact parameters, which correspond to all points of the image plane where the observer is located. In Section \ref{sec:SIE}, we describe the numerical scheme to integrate in time. Particularly, we carried out the simulations with four different adaptive time step time integrators and show the norm of the Hamiltonian constraint for each one of them. Besides, as an example, we simulate the shadow produced by the Kerr black hole. In particular, we compare the numerical solution obtained with \texttt{OSIRIS} with the respective analytical solution for each time integrator. Here we show the capability of the code to deal with several initial conditions. Subsequently, to compute the gravitational lensing, it is necessary to classify the orbits to elucidate the effect of gravity over the light. For this reason, in Section \ref{sec:CS}, we describe this phenomenon and its implementation in \texttt{OSIRIS}. In Section \ref{sec:TAD}, we show the shadow produced by the Schwarzschild black hole surrounded by a Thin accretion disk. Finally, in Section \ref{sec:CON}, we present our main conclusions and discussions.

Throughout this paper, we employ the signature $(-,+,+,+)$ and geometrized units, in which the gravitational constant and the speed of light are equal to unity. Additionally, we set the mass of the black hole $M_{BH}=1$. 

\section{Shadow formulation}\label{sec:meq}
\subsection{Motion equations} 

The main goal of \texttt{OSIRIS} is to evolve null geodesics in asymptotically flat and stationary axisymmetric space-times, which implies that there are two constants of motion: the energy $E$ and the azimuthal angular momentum $L$. In this way, given a metric tensor 

\begin{equation}
\textbf{g} = g_{\alpha\beta}dx^\alpha\otimes dx^\beta, \label{eq:le}
\end{equation} 
being $g_{\alpha\beta}$ the contravariant components of the tensor and $dx^\alpha$ the base of 1-forms, we can obtain the motion equations from the hamiltonian formulation
\begin{equation}
\dot{p}_{\mu} = - \frac{\partial H}{\partial x^{\mu}} \; \; , \; \; 
\dot{x}^\mu = \frac{\partial H}{\partial p_{\mu}}, \label{eq:edm}
\end{equation}
where the overdot implies differentiation with respect to an affine parameter $\lambda$. The Hamiltonian is defined in terms of the contravariant components of the metric tensor, $g^{\alpha\beta}$, and for null particles, it satisfies that
\begin{equation}
H = \frac{1}{2} g^{\mu\nu} p_\mu p_\nu = 0, \label{eq:ligH}
\end{equation}
where $p_\alpha$ are the components of the four-momentum.  For simplicity and without loss of generality, we will choose the labels $\{x^\mu\} = \{t,r,\theta,\phi\}$ for the coordinate system to be spherical-like. It is worth mentioning that  \texttt{OSIRIS} also evolves time-like geodesics. In \ref{appendix:A}, we show trajectories for test particles around a compact object with a non-vanishing quadrupole moment. Thus, from the expressions in  \ref{eq:edm} and \ref{eq:ligH}, it is possible to obtain the coordinates and momentum of the photons' trajectories as follows
\begin{widetext}
$$
\dot{t} = g^{tt}p_{t} + g^{t\phi}p_{\phi}, \qquad
\dot{r} = g^{rr}p_{r}, \qquad
\dot{\theta} = g^{\theta \theta}p_{\theta}, \qquad
\dot{\phi} = g^{t\phi}p_{t} + g^{\phi \phi}p_{\phi}, 
$$
$$
\dot{p}_{t} = 0,
$$
$$
\dot{p}_{\phi} = 0,
$$
$$
\dot{p}_{r} = \frac{-p_{t}^{2} \partial_{r}g^{tt} -2p_{t}p_{\phi}\partial_{r}g^{t\phi} -p_{r}^{2} \partial_{r}g^{rr} -p_{\theta}^{2} \partial_{r}g^{\theta \theta} -p_{\phi}^{2} \partial_{r}g^{\phi \phi}}{2},
$$
$$
\dot{p}_{\theta} = \frac{-p_{t}^{2} \partial_{\theta }g^{tt} -p_{r}^{2} \partial_{\theta}g^{rr} -p_{\theta}^{2} \partial_{\theta}g^{\theta \theta} -p_{\phi}(2p_{t}\partial_{\theta}g^{t\phi} + p_{\phi} \partial_{\theta}g^{\phi \phi})}{2},
$$
\end{widetext}

those are the expressions that we solved with \texttt{OSIRIS}.
\subsection{Backward ray-tracing}\label{sec:BRT}
\begin{figure}[htbp]
 \centering
 \includegraphics[width=0.48\textwidth]{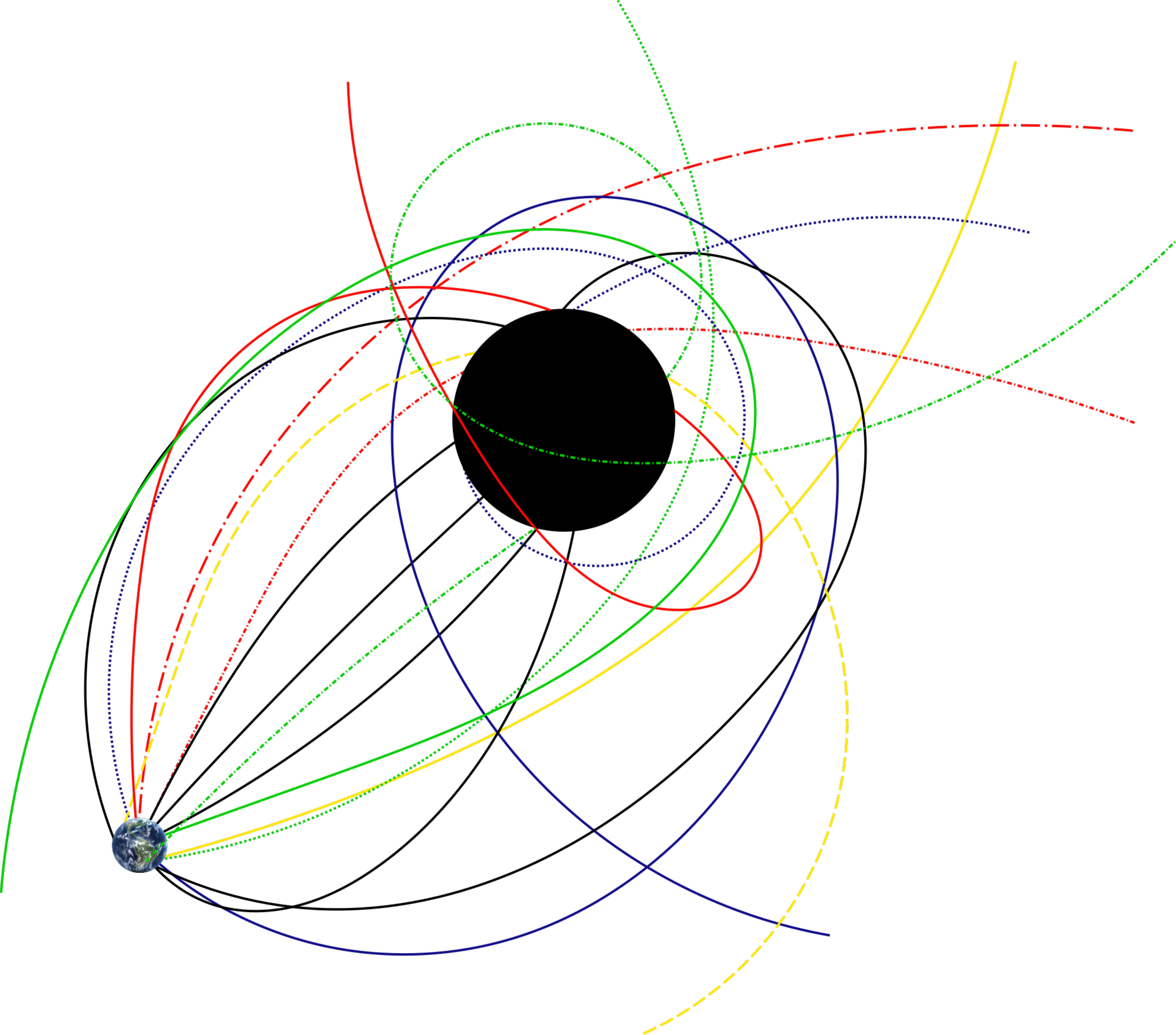}
 \caption[Backward ray tracing graphic representation]{Graphic representation of the backward ray tracing: the Earth is the observer from which the photons are emitted, and the black sphere plays the role of a black hole. The lines represent null geodesics, among which are those that are trapped by gravitational attraction (black lines) and those that escape to infinity (colored lines).}
 \label{fig:BRT}
\end{figure}

To visualize the effects of the strong gravitational field produced by a black hole, it is necessary to analyze the trajectories followed by photons that orbit in its vicinity and translate the information of those that escape and manage to be detected. In essence, this is the main algorithm of \texttt{OSIRIS}, and to understand its operation, we propose two scenarios:
\\ \\
\textbf{Scenario 1:} a bright source emits photons in all directions. Among all, it is possible to find those that orbit around the black hole and escape to infinity, others that fall into the event horizon, or even those that do not approach the black hole. The photons that escape and arrive at the observer represent a minimal fraction of all the radiation originally emitted by the source, whereby calculating numerically the motion equations in this scenario translates into a waste of computational time. In other words, this scenario is not optimal and therefore is discarded.
\\ \\
\textbf{Scenario 2:} The observer is the origin of the null geodesics that evolve back in time.  In this way, it is possible to reconstruct the bright source from which they come. This method is known as backward ray tracing, and it is a standard in the simulation of numerical shadows \citep{vincent2011gyoto}. In figure \ref{fig:BRT}, we illustrate this method, where the gravitational source is a black hole, and the trajectories followed by photons that fall into the event horizon are classified with black color. Whereby, those are physically undetected geodesics whose starting point is the observer. 
%
%
\subsection{Impact parameters} \label{sec:IP}
\begin{figure}[htbp]
\centering
\includegraphics[scale=0.056]{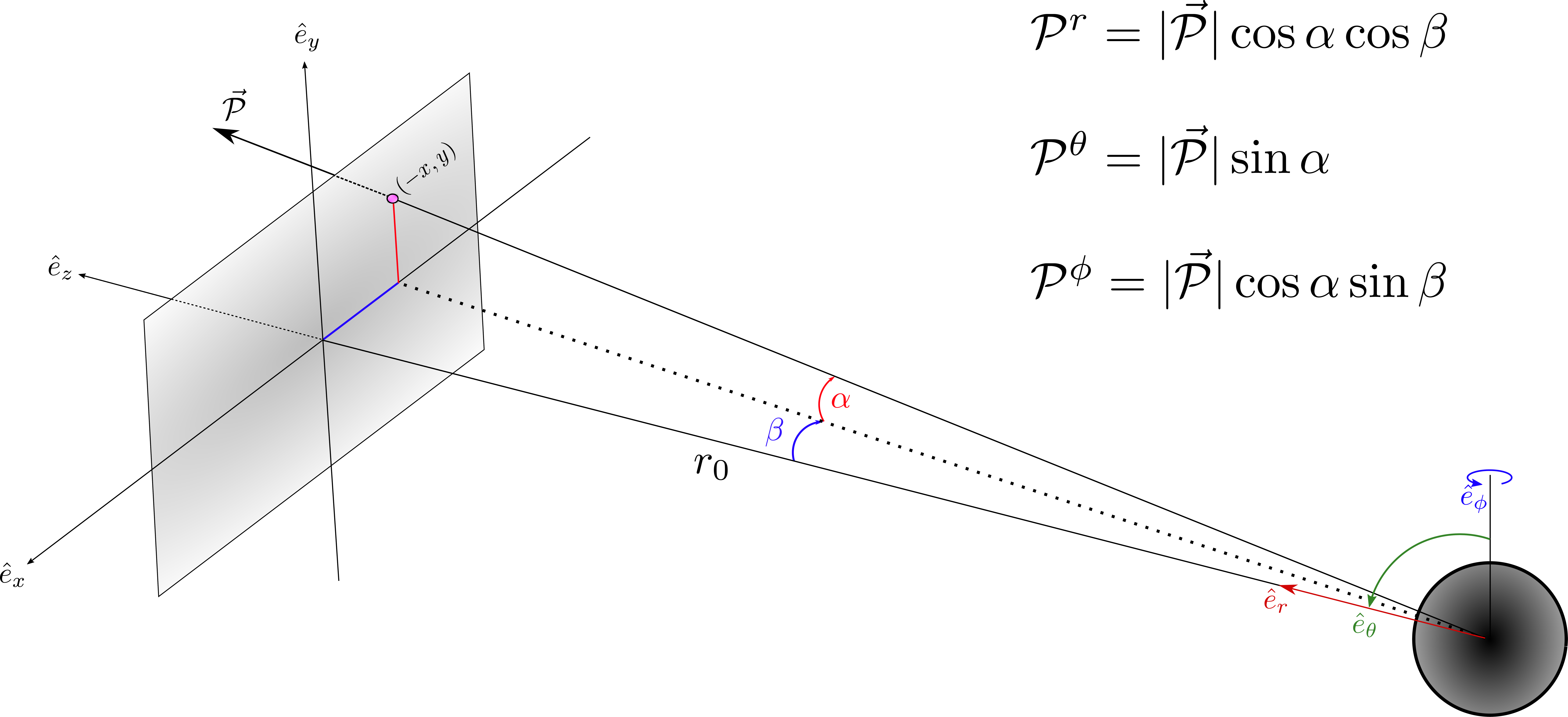}
\caption[Impact parameters]{Illustration of the image plane associated with an inertial observer at a distance $r_{0}$ from the black hole. The final position of the photons is characterized by the spatial part of four-momentum $\vec{\mathcal{P}}$ measured by the observer (the temporal component is irrelevant since it does not influence the trajectory of the photons), whose origin is taken in the center of the black hole. The impact parameters are related to $\alpha$ and $\beta$, where $\alpha$ is the angle between $\vec{\mathcal{P}}$ and its projection in the plane $xz$, and $\beta$ corresponds to the angle between this projection and $r_{0}$. Also,  the observer is located such that the radial direction coincides with the $z$-axis, where it is clear that $\hat{e}_x = \hat{e}_\phi$ and $\hat{e}_y = - \hat{e}_\theta$. }\label{fig:momentums}
\end{figure}

To obtain the numerical shadow produced by a black hole, it is necessary to know the information encoded in the photons that orbit in its vicinity. It is then defined as impact parameters to all points $(x, y)$ of the image plane where an observer, located at infinity, makes the measurements \cite{Johannsen_2013}, as we illustrate in figure \ref{fig:momentums}.

Since the observer is far from the source, it is possible to make a small angles approximation
\begin{equation}
x = -\beta r_{0}, \quad y = \alpha r_{0}, \label{eq:pim}
\end{equation}
and
\begin{equation}
\mathcal{P}^{r} = \big|\mathcal{\vec{P}}\big|, \quad \mathcal{P}^{\theta} = \alpha \big|\mathcal{\vec{P}}\big|, \quad \mathcal{P}^{\phi} = \beta \big|\mathcal{\vec{P}}\big|, \label{eq:pcomp}
\end{equation}
 where $\mathcal{P}^{\alpha}$ correspond to the components of the four-momentum in the observer's frame. These are determined from the canonical momentums through the transformation
\begin{equation}
\mathcal{P}^{\alpha} = \eta ^{\alpha \mu} {\Lambda _\mu}^{\nu} p_{\nu},
\end{equation}
where $\eta ^{\alpha \mu}$ are the components of the metric tensor in the Minkowski space-time, and ${\Lambda _\mu}^{\nu}$ are the components of a base change matrix that relates the momentums measured in both reference systems, which is given by
\begin{equation}
[{\Lambda _{\mu}}^{\nu}] = \left(\begin{array}{cccc}
A_{t} & 0  &  0 &  \;\; -A_{t} \frac{g_{t\phi}}{g_{\phi \phi}} \\ 
0  & \frac{1}{\sqrt{ g_{rr} }} &  0 & \;\; 0  \\
0  & \;\; 0 & \frac{1}{\sqrt{ g_{\theta \theta} }} &  \;\; 0    \\
0 & \; \;  0 & 0 & \;\; \frac{1}{\sqrt{ g_{\phi \phi} }}
\end{array}\right), \label{eq:matrix}
\,\,
A_{t} = \sqrt{ \frac{g_{\phi \phi}}{g^{2}_{t \phi} - g_{tt} g_{\phi \phi}}  }. \nonumber
\end{equation}
Explicitly, the momentums can be written as follows
\begin{gather}
\mathcal{P}^{t} = A_{t} \left[p_{t} +  \frac{g_{t\phi}}{g_{\phi \phi}} L \right], \nonumber
\quad
\mathcal{P}^{r} = \frac{p _{r}}{\sqrt{g_{rr}}},
\\ \label{eq:momentums} \\ 
\mathcal{P}^{\theta} = \frac{p _{\theta}}{\sqrt{g_{\theta \theta}}}, \quad
\mathcal{P}^{\phi} = \frac{L}{\sqrt{g_{\phi \phi}}}. \nonumber
\end{gather}
Additionally, the magnitude of the four-momentum satisfies the Hamiltonian constrain
\begin{equation}
\big|\mathcal{P} \big|^{2} = -(\mathcal{P}^{t})^{2} + (\mathcal{P}^{r})^{2} + (\mathcal{P}^{\theta})^{2} + (\mathcal{P}^{\phi})^{2} = 0. \label{eq:const}
\end{equation}
Therefore, the expressions in \ref{eq:pcomp} take the form
\begin{equation}
\alpha = \frac{\mathcal{P}^{\theta}}{\mathcal{P}^{t}},
\quad \beta = \frac{\mathcal{P}^{\phi}}{\mathcal{P}^{t}}, \label{eq:ab}
\end{equation}
and replacing \ref{eq:ab} in \ref{eq:pim}, we obtain
\begin{equation}
x = -r_{0} \frac{\mathcal{P}^{\phi}}{\mathcal{P}^{t}}, \qquad y = r_{0} \frac{\mathcal{P}^{\theta}}{\mathcal{P}^{t}}. \label{eq:pimf}
\end{equation}
The equations in \ref{eq:pimf} are general expressions for the impact parameters in an axisymmetric background, which only depends on the metric tensor components \cite{Pedro}. Thus, setting set $\mathcal{P}^t = \mathcal{E} = 1$ without loss of generality, and combining equations \ref{eq:pimf} and \ref{eq:momentums}, initial momentums can be computed in terms of each point on the image plane as follows

\begin{gather}
p_{\theta} = \frac{y\sqrt{g_{\theta\theta}}}{r_0}, \nonumber
\quad
E = \frac{1}{A_t} + \frac{x g_{t\phi}}{r_0\sqrt{g_{\phi\phi}}},
\\ \label{eq:momentums} \\ 
L = -\frac{x\sqrt{g_{\phi\phi}}}{r_0}, \quad
p_r = \sqrt{g_{rr}\left [ 1 - \left( \mathcal{P}^{\theta}\right)^2  - \left( \mathcal{P}^{\phi}\right)^2 \right ]}. \nonumber
\end{gather}
This image plane setup has been used to numerically calculate shadows generated by compact objects in various space-times, as well as to study the impact of the intense gavitational field on phenomena such as the so-called gravitational lensing \cite{Letter, Pedro, Vincent_2016, cunha_2016, Wang, Wang_2018}. Furthermore, thanks to this configuration, in some particular cases in which the system of equations is variable-separable, it is possible to find analytical solutions to determine the rim of the shadow through the calculation of the spherical orbits of photons around certain compact bodies. Next, we show an example in Kerr space-time.
%
\section{Orbits in curved space$-$time} \label{sec:SIE}
\subsection{Single orbits around Kerr black hole} 
In order to solve the set of equations, let us first describe the criteria used to choose the best numerical integrator for our purpose. In general, four different integration schemes with adaptive step have been tested: Runge-Kutta Fehlberg 45 (RKF45) \cite{RKF_1969}, Runge-Kutta Cash-Karp 45 (RKCK45) \cite{Cash_Karp_1990}, Runge-Kutta Dormand-Prince 45 (RKDP45) \cite{RKDP_1980} and a Bulirsch-Stoer (BS) algorithm \cite{BS_1966}. The Runge-Kutta methods are the commonly staged methods used for the approximation of solutions of ordinary differential equations. On the other hand, the BS algorithm combines three different ideas: Richardson extrapolation that considers the final answer of a numerical calculation as an analytic function of an adjustable parameter like the stepsize. Once there is enough information about the function, it is fit to some analytic form through a rational function extrapolation, to finally carry out the integration by the modified midpoint method \cite{Num_Rec_1992}.\\

\begin{figure}[htbp]
\includegraphics[scale=0.135]{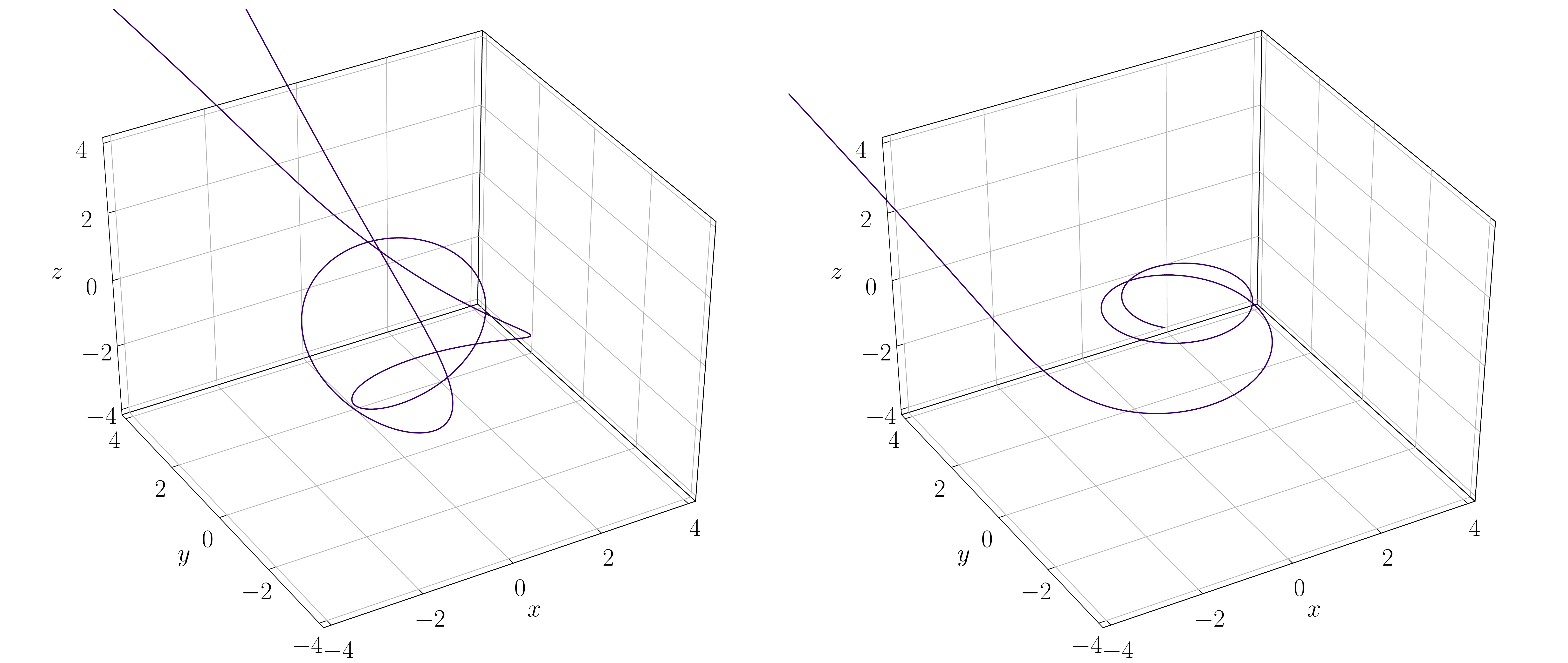}
\caption[Orbits around Kerr black hole]{Orbits around a Kerr black hole with spin parameter $a=0.98$. Left panel: escape orbit; right panel: falling orbit.} \label{fg:orbits}
\end{figure}

Preserving the Hamiltonian constraint for a photon that orbits around a Kerr black hole with dimensionless spin parameter $a = 0.98$ and initial conditions $(t_0,r_0, \theta_0, \phi_0)$, we considered two scenarios: \begin{itemize}
\item[i] escaping from the gravitational attraction (figure \ref{fg:orbits}, left panel),
\item[ii] falling into the event horizon (Figure \ref{fg:orbits}, right panel),
\end{itemize}
 according to the hamiltonian constraint in \ref{eq:ligH}, such that
\begin{equation}
\text{Error} = |H_{num} - H| = |H_{num}|.
\end{equation}

\begin{figure}[h]
\includegraphics[width=0.4\textwidth]{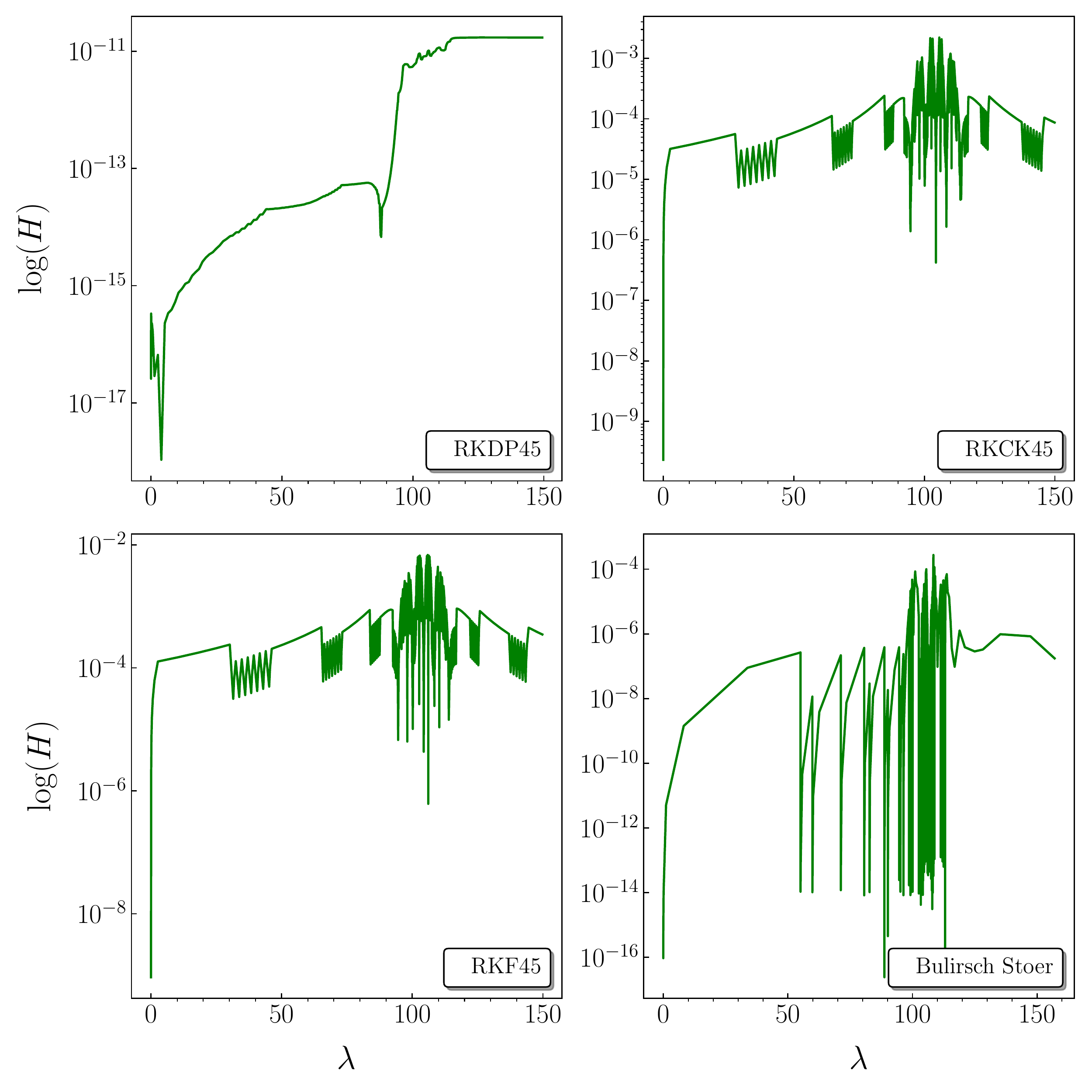}
\caption[Hamiltonian constrain for a photon that escapes]{Hamiltonian constraint in logarithmic scale for a photon that escapes. Upper left panel: RK Dormand-Prince; upper right panel: RK Cash-Karp; lower left panel: RK Fehlberg; lower right panel: BS algorithm. The initial conditions are: $r_0 = 100$, $\theta_0 \approx 1.570796$, $\phi_0 \approx 1.570796$, $p_t \approx -0.989953$, $p_r \approx -1.009723$, $p_{\theta} \approx 1.870000$ and $p_{\phi} \approx 2.000098$.} \label{fg:scapes}
\end{figure}

\begin{figure}[htbp]
\includegraphics[width=0.4\textwidth]{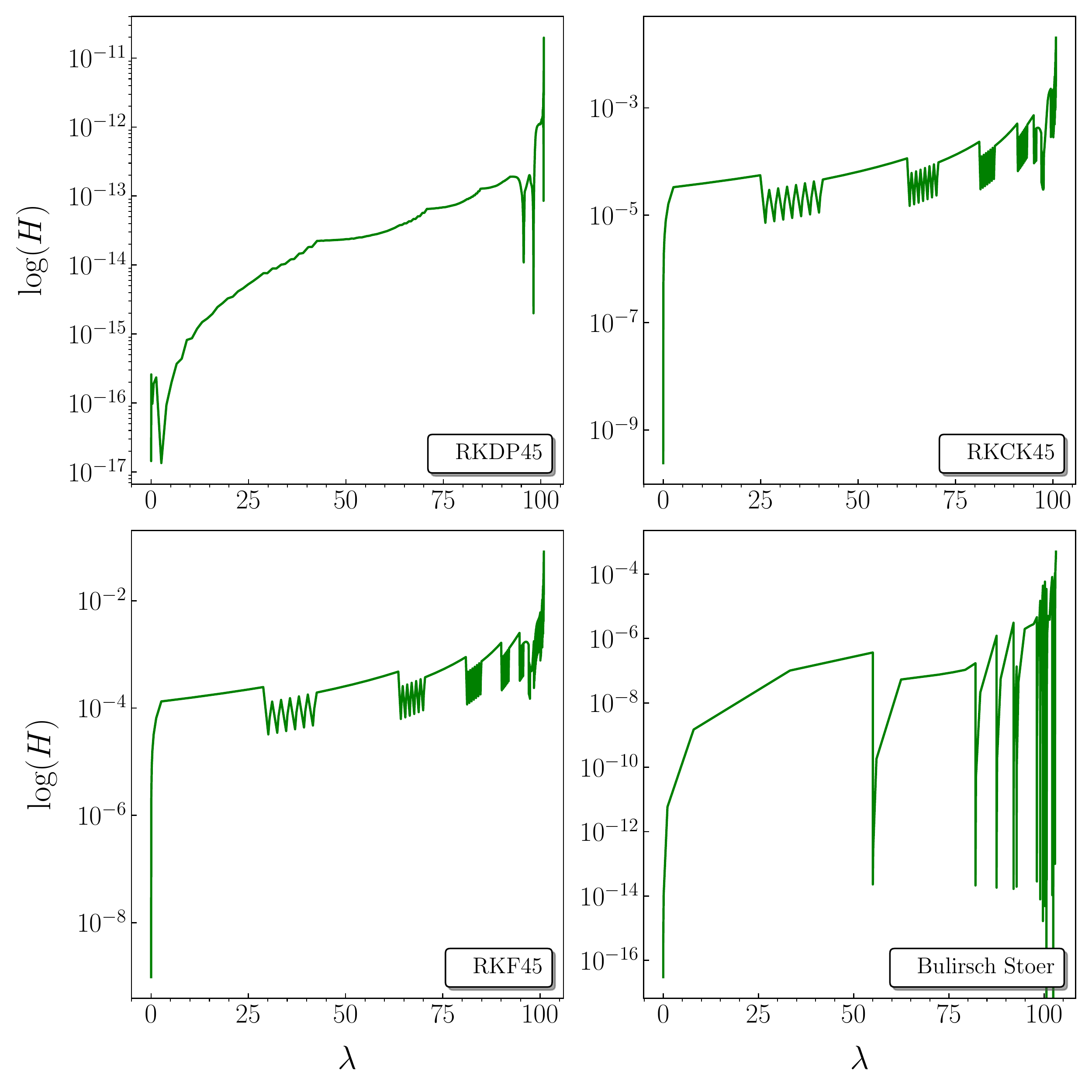}
\caption[Hamiltonian constrain for a photon that falls]{Hamiltonian constraint in logarithmic scale for a photon that falls. Upper left panel: RK Dormand-Prince; upper right panel: RK Cash-Karp; lower left panel: RK Fehlberg; lower right panel: BS algorithm. The initial conditions are: $r_0 = 100$, $\theta_0 \approx 1.570796$, $\phi_0 \approx 1.570796$, $p_t \approx -0.989952$, $p_r \approx -1.009314$, $p_{\theta} \approx 3.750000$ and $p_{\phi} \approx 1.250061$.} \label{fg:falls}
\end{figure}

In figures \ref{fg:scapes} and \ref{fg:falls} we plot the hamiltonian norm for two particular orbits: an escape one and a fall one, respectively. Our numerical results suggest that in both cases, the RK Dormand-Prince preserves the hamiltonian norm better than one part in $10^{-10}$, being the best option, with the BS algorithm the second one. It is worth mentioning that Dormand-Prince method has seven-time steps but only uses six function evaluations per step. This method chooses the coefficients to minimize the error of the fifth-order solution. It is the main difference with the Fehlberg Runge-Kutta time integrator, which was constructed so that the fourth-order solution has a small error. It is clear that the error grows as the photon approaches the event horizon at $r_0 = 100$. The oscillations in the curves with each integrator occur due to the step refinement in regions where the gravitational field is stronger, i.e., near to the event horizon.  \\

%
\subsection{Black hole shadow in Kerr space$-$time} \label{sec:BHSK}

\begin{figure}[h!]
\centering
\includegraphics[width=0.45\textwidth]{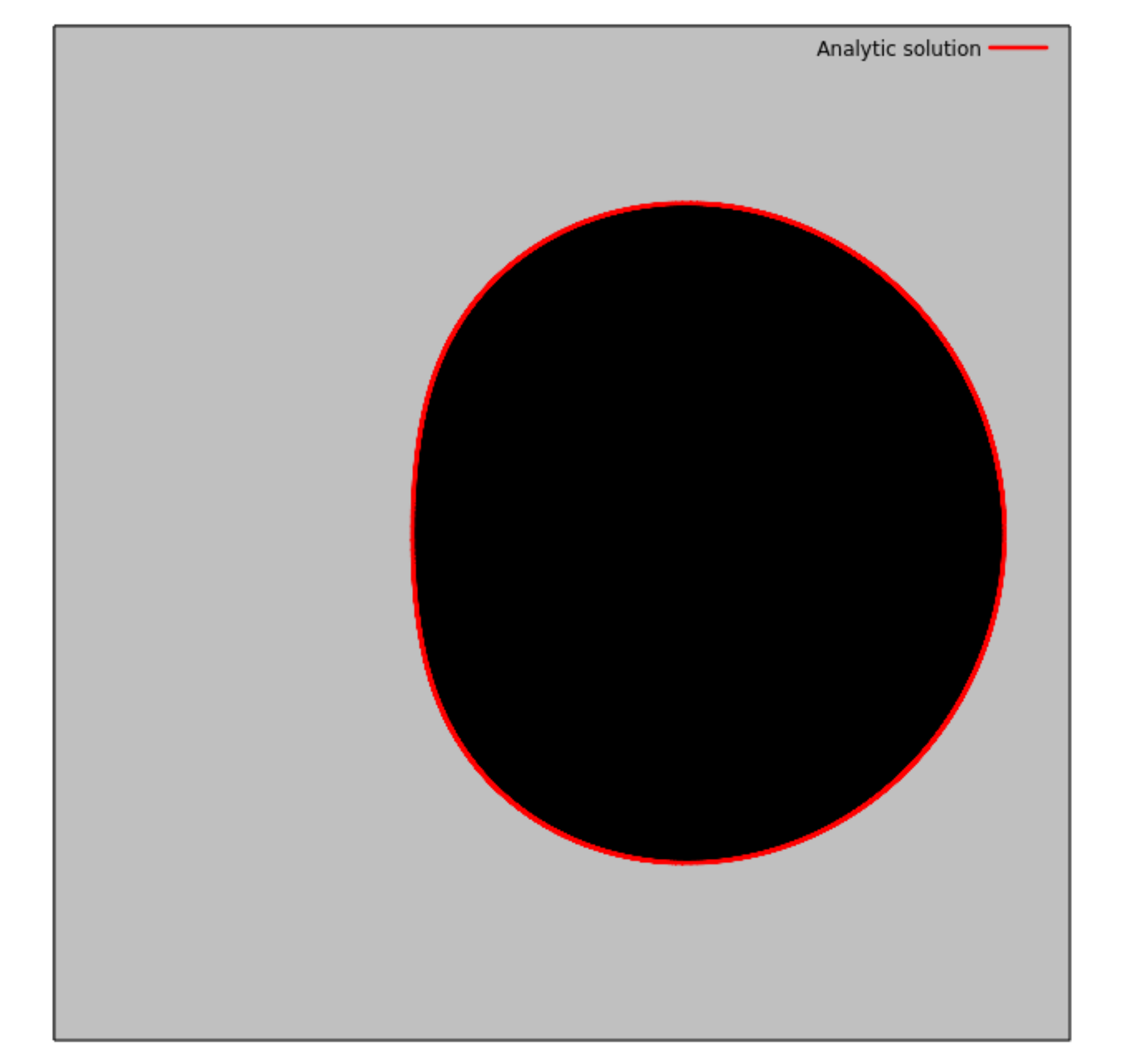}
\caption[Shadow]{Numerical (black area) and analytical (red rim) black hole shadow in the Kerr space-time with a spin parameter $a = 0.98$ for an observer located in the equatorial plane, $\theta_0 = \pi/2$ at $r_0 = 1000$. The domain of the image plane is $-8\leq x,y \leq 8$. This result is in agreement with the obtained by \cite{pu2016odyssey}, but with a dimensionless spin parameter $a = 0.998$.} \label{fg:shadow}
\end{figure}

\begin{figure}[htbp]
\centering
\includegraphics[width=0.47\textwidth]{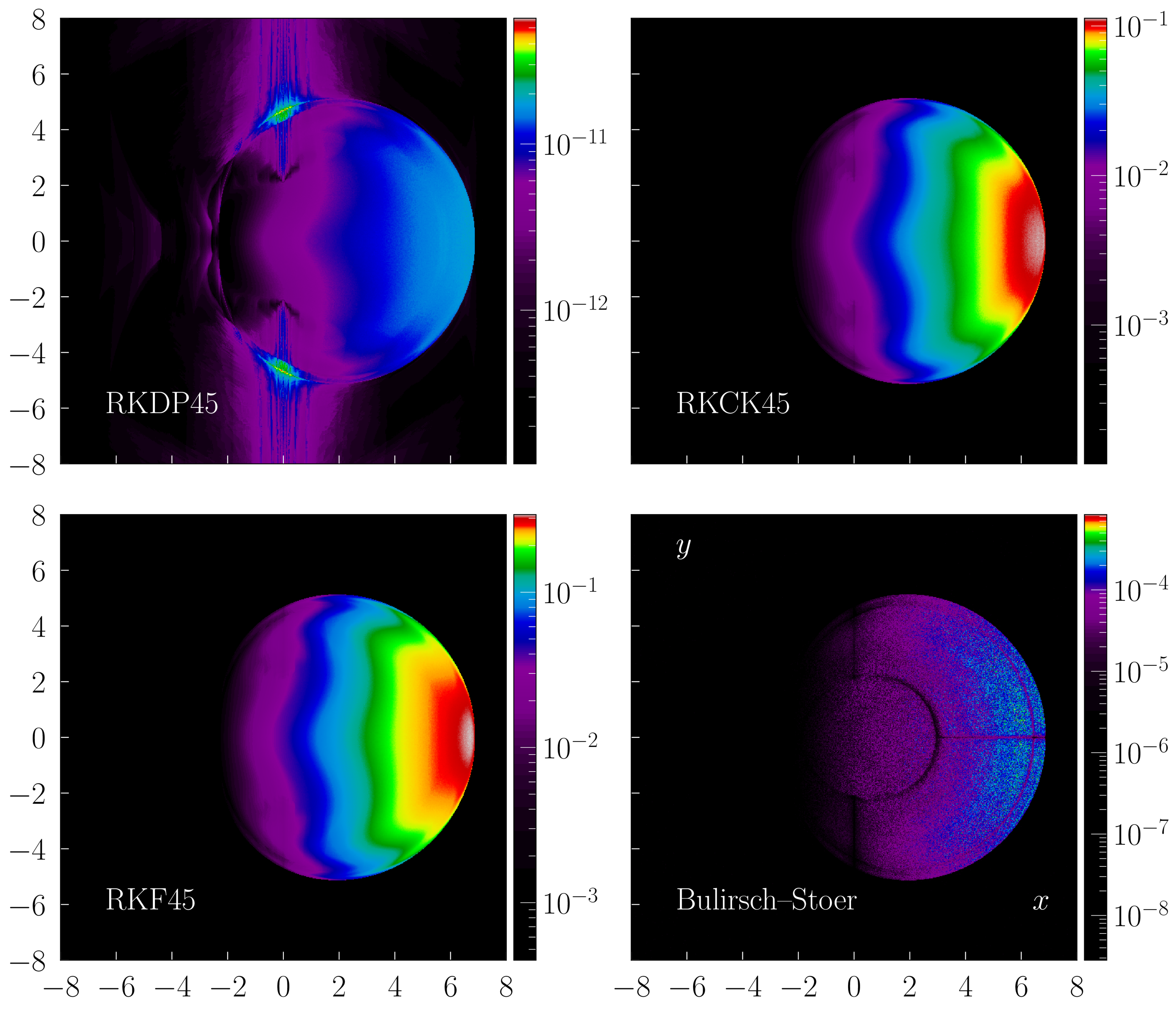}
\caption[Average of of the hamiltonian norm]{Average of the hamiltonian constraint with $r_0 = 100$. Upper left panel: RK Dormand-Prince; upper right panel: RK Cash-Karp; lower left panel: RK Fehlberg; lower right panel: BS algorithm} \label{fg:average_norm}
\end{figure}

\begin{figure}[htbp]
\centering
\includegraphics[scale=0.4]{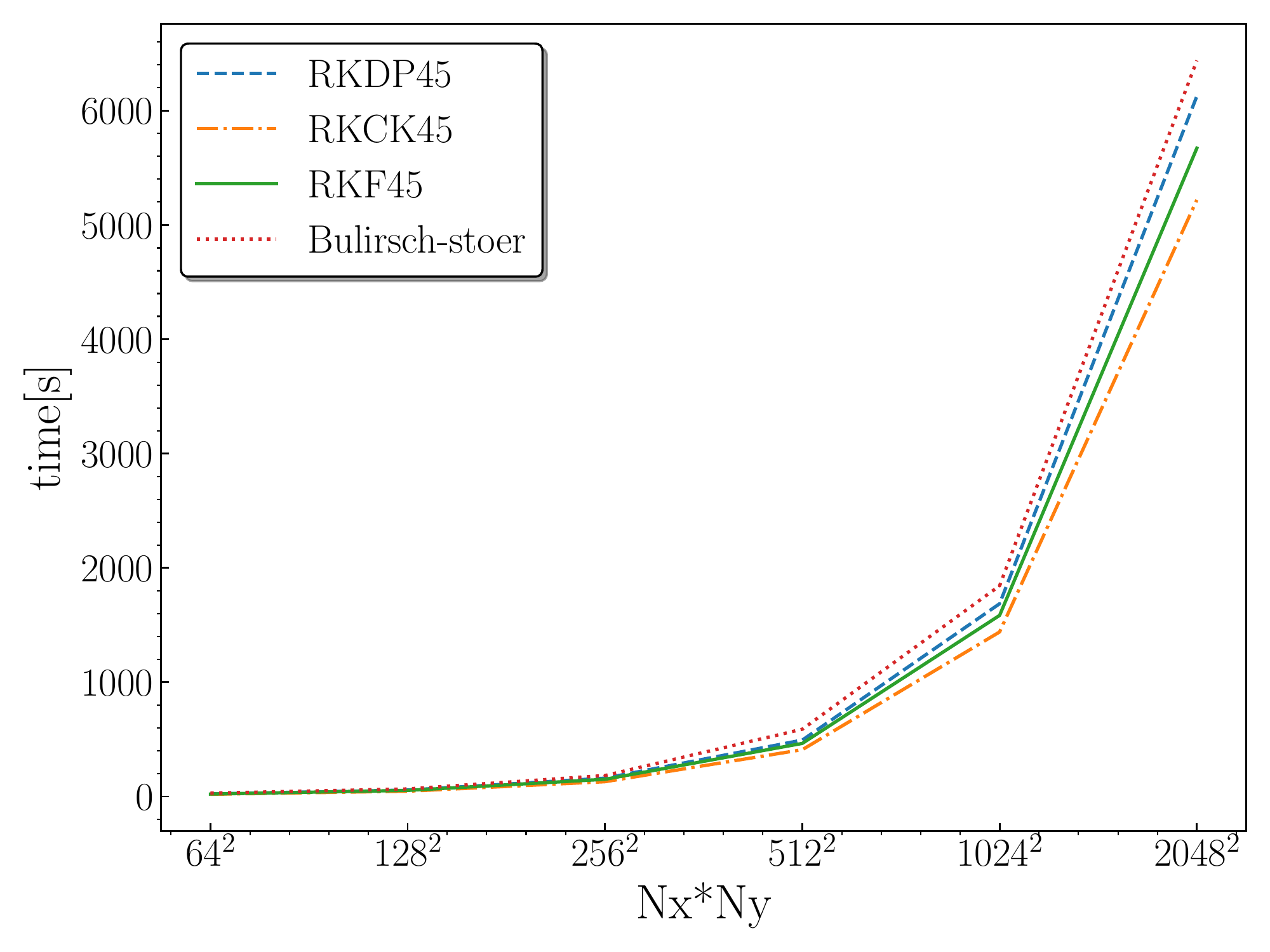}
\caption[Computational time]{Computational time (time machine in seconds) for all the integration schemes as a function of the numerical resolution.} \label{fg:comp_time}
\end{figure}

The analytical solution for the shadow of a Kerr black hole is given by \cite{florov}
\begin{gather}
x = -\frac{\xi}{\sin \theta}, \qquad y = \pm \sqrt{\eta - \cos ^{2}\theta \left( \frac{\xi ^{2}}{\sin ^{2} \theta} - a^{2} \right)},  \label{eq:anal}
\end{gather}
 where the points $(x, y)$ are parameterized by the radii $r$ of the spherical orbits through $\xi$ and $\eta$, which are given by the expressions
 \begin{align}
\xi    =& - \frac{r^{3} -3Mr^2 + a^{2}r + a^{2}M}{a (r - M)}, \nonumber \\ \\
\eta =& - \frac{r^{3} (r^{3} - 6Mr^{2} + 9M^{2}r - 4a^{2}M)}{a^{2} (r - M)^2}. \nonumber \label{eq:xie}
\end{align}
 In figure \eqref{fg:shadow}, we show the analytical and numerical shadow of a black hole with a spin parameter $a=0.98$. Comparing both results, we obtained that the simulation is a little smaller than the analytical. It is because this solution is calculated at infinity, but it is not possible to simulate an observer with these characteristics; in addition, initializing the motion too much larger radii means that the error accumulates more and more between iterations.
On the other hand, to carry out a global analysis, we plot the average of the hamiltonian norm in a square scattering region bounded by $-8 \leq x \leq 8$ and $-8 \leq y \leq 8$, with a dense uniform grid of $625 \times 625$ initial conditions. Using a logarithmic color palette, we displayed the final value of the hamiltonian norm for the entire phase space (Figure \ref{fg:average_norm}). As we have seen before, the RK Dormand-Prince preserves the constraint better than a part in $10^{-11}$ (in the worst scenario e.g., near the black hole, better than a part in $10^{-10}$). In contrast, the remaining methods “fail” to preserve the constraint, yielding, in the best case, to values of the order of $10^{-3}$ for the RK methods, and of the order of $10^{-8}$ in the case of the BS algorithm. 
\\
Finally, Figure \ref{fg:comp_time} exposes simulation time machine as a function of the mesh resolution. We can see that the differences between the times required by each integrator increase with the resolution. Thus, at higher resolution, the RK Dormand-Prince and the BS algorithm require more computational time than the RK Cash-Karp and the RK-Fehlberg. However, due to the better numerical behavior of the former when maintaining the norm, and the almost negligible increase in the time machine, the RK-Dormand-Prince emerges as the best option for our purpose. Based on the above, we compare our result with another codes. In \cite{chan2013gray}, it is possible to appreciate a time comparison between three different codes: \texttt{GRay} \cite{chan2013gray}, \texttt{GeoKerr} \cite{dexter} and  \texttt{Ray} \cite{Dimitrios}, obtaining that in the process of integration of $10^6$ geodesics the time orders in seconds employed for each code respectively were $10^3$, $10^4$ and $10^5$, making of \texttt{GRay} the fastest of the three. \texttt{OSIRIS}'s algorithm evolves $1024^2$ geodesics in the same order of time as \texttt{GRay}, see Figure \ref{fg:comp_time}. Furthermore, we show that \texttt{OSIRIS}, besides being fast, preserves both the the hamiltonian constrain and the hamiltonian norm below $10^{-11}$. Nevertheless, comparing our results with \texttt{Odyssey} \cite{pu2016odyssey}, it evolves $1024^2$ geodesics in two orders of magnitude of time less than \texttt{OSIRIS}. It is necessary to mention that both \texttt{GRay} and \texttt{Odyssey} employ a powerful programming architecture based in GPU usage and parallelization in CUDA, in which each code evolves millions of geodesics at the same time. 
\\
With the error analysis discussed in this section, we concluded that the calibration of the code is optimal. Then, we proceed to study the influence of the gravitational field on the trajectory of the photons.
%
\section{Relativistic effects}\label{sec:CS}
\subsection{Celestial sphere} 

\begin{figure}[htbp]
\centering
\includegraphics[scale=0.25]{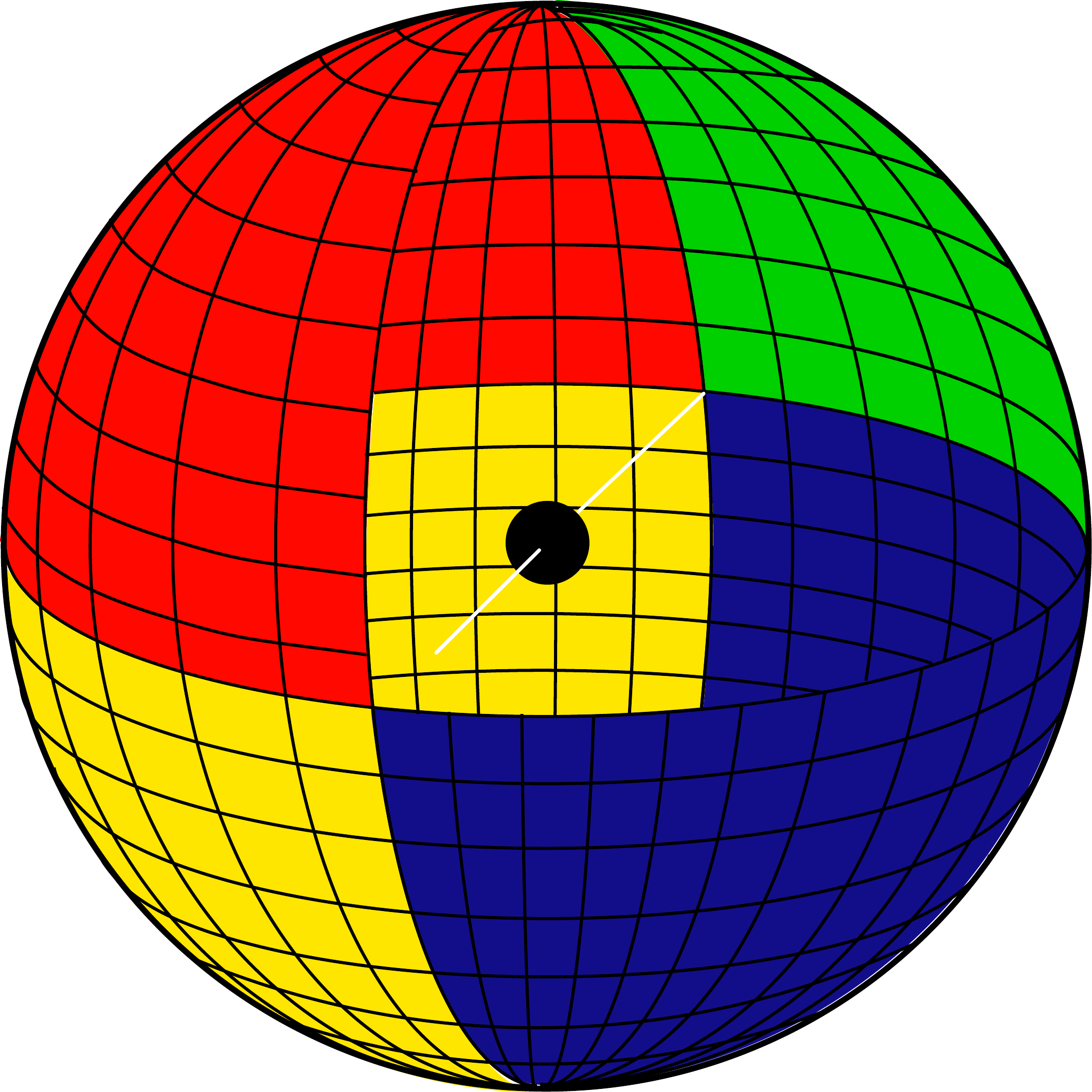}
\caption[Celestial sphere scheme]{Scheme of the celestial sphere where we remove a section of this to see inside the observer (Earth) and the compact object. The sphere is divided into four quadrants as follows.
\textbf{Top}, for $0 \leq \theta < \pi/2$: green quadrant, if $0 \leq \phi < \pi$; red quadrant, $\pi \leq \phi < 2\pi$. \textbf{Bottom}, for $\pi/2 \leq \theta < \pi$: blue quadrant, if $0 \leq \phi < \pi$; yellow quadrant, if $\pi \leq \phi < 2\pi$.
The white line represents the direction of observation that coincide with radial direction.} \label{fg:sph}
\end{figure}

Now we are going to classify the orbits to elucidate the effect of gravity on null geodesics. For this purpose, following \cite{Bohn_2015}, we define a celestial sphere as a bright source from which the light rays will be emitted. This sphere is concentric with the black hole, surrounds the observer, and is divided in four colors: blue, green, yellow, and red. Furthermore, the notion of curvature is provided by a black mesh with constant latitude and longitude lines, separated by $6^\circ$ (figure \ref{fg:sph}). In this way, the classification will be as follows: if the photon's radial position, $r$, satisfies the condition $r < r_{H} + \delta _{r}$, where $ r_{H} $ is the event horizon radius and $ \delta _{r} $ is a little buffer, then it corresponds to a black point in the image plane. On the contrary, if the photon escapes from the strong gravitational attraction and $ r>r_{cs} $, where $ r_{cs} $ is the radius of the celestial sphere, the integration process is stopped and a color is assigned depending on the sector of the sphere where the photon strikes.

\subsection{Gravitational lensing} \label{sec:GL}

\begin{figure}[htbp]
\centering
\includegraphics[scale=0.4]{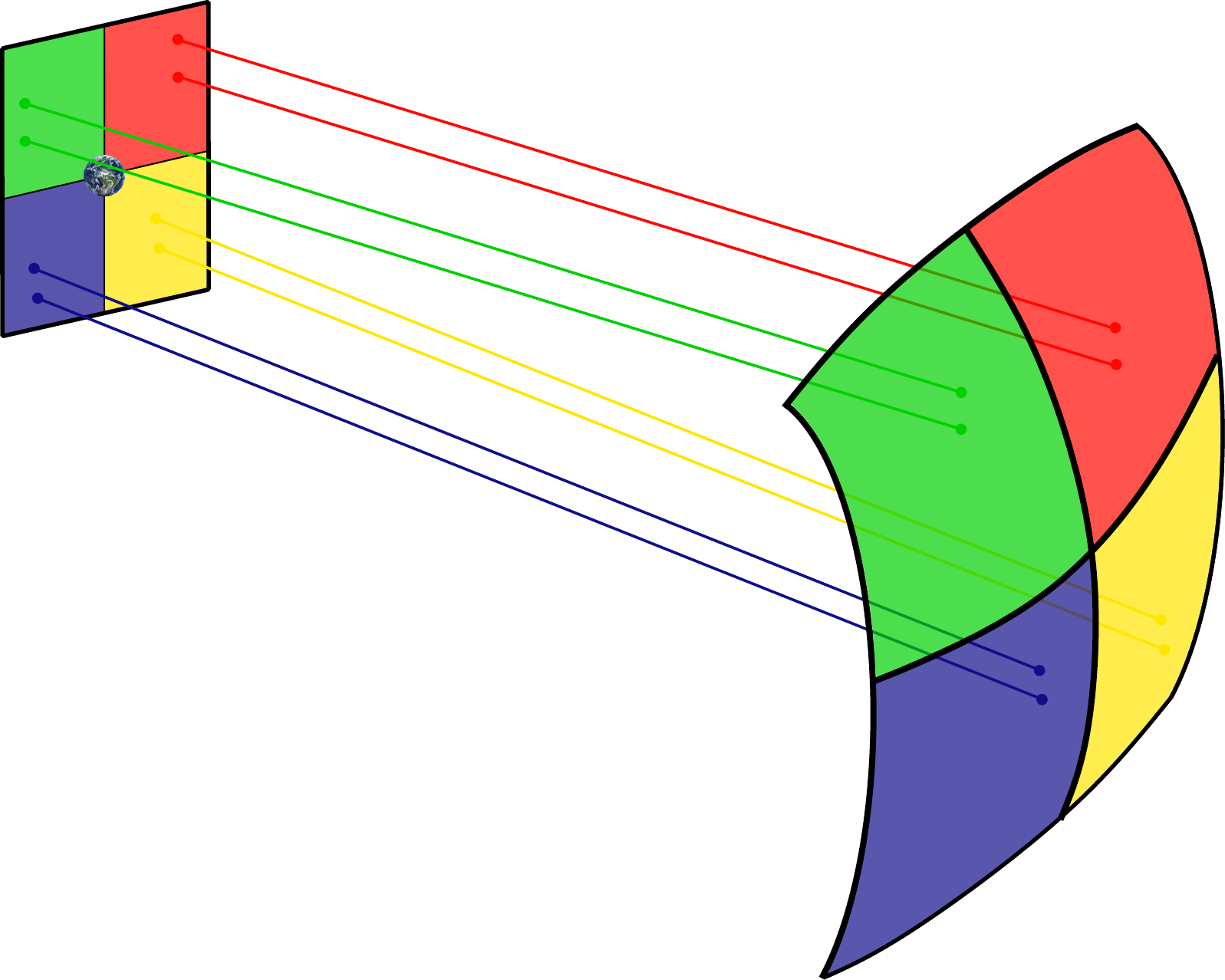}
\caption{Graphic representation of the trajectory of photons in the Minkowski space-time. The observer is looking directly at the celestial sphere, and in the absence of any gravitational field source, the path of the photons is a straight line. Therefore, in the observer's plane, we obtain an image that corresponds with a section of the celestial sphere.} \label{fg:ms}
\end{figure}

\begin{figure}[htbp]
\centering
\includegraphics[scale=0.085]{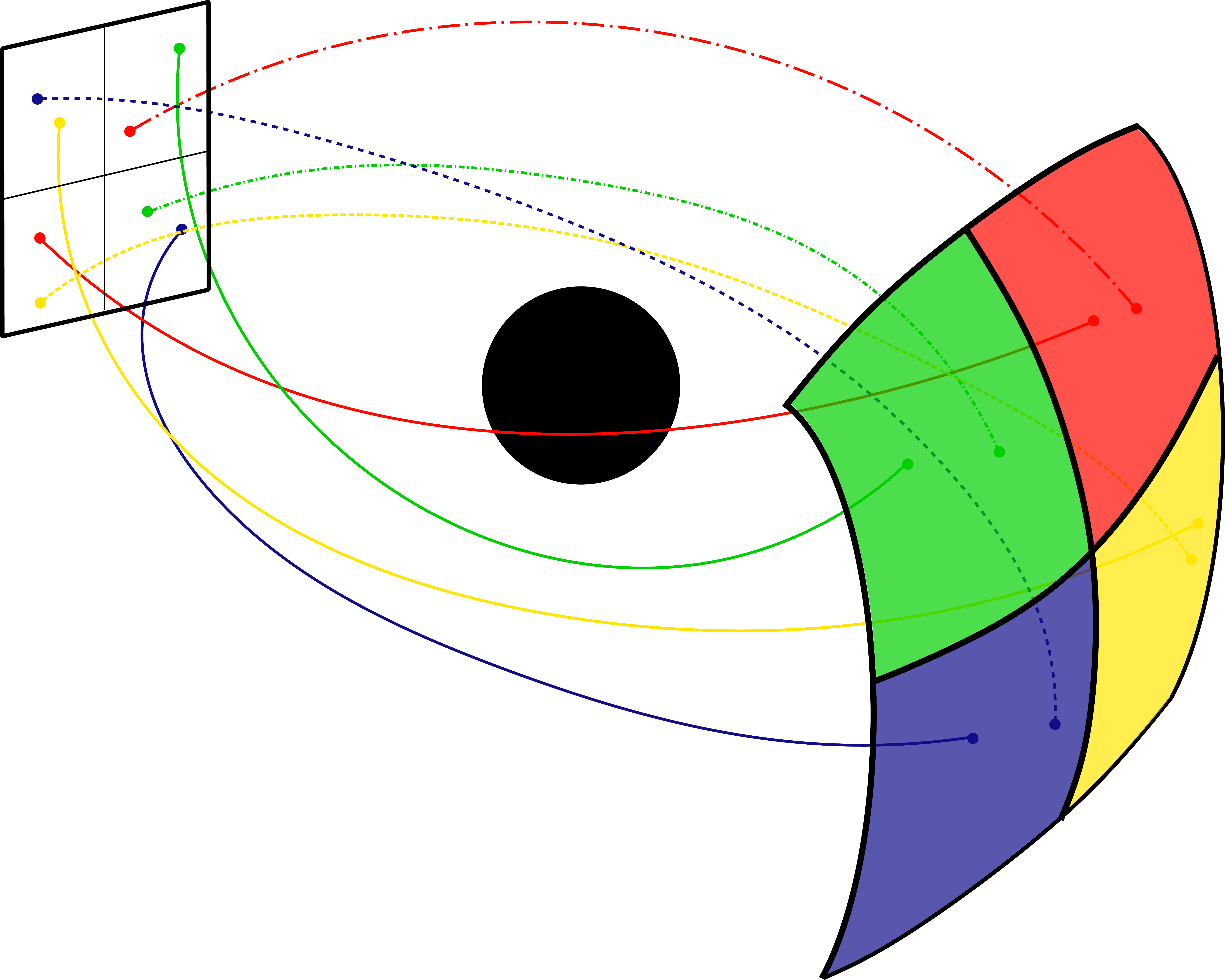} \caption{Graphic representation of the deflection in the path of the photons in the presence of a compact object. Due to the strong gravitational field, some photons emitted from a quadrant can be detected in the image plane as if they had a different point of origin.} \label{fg:gls}
\end{figure}

Due to the deflection of light in the presence of a compact object, the optical perception that we could obtain from a bright source can provide interesting results at the time of observation. This phenomenon is known as gravitational lensing. To have a better idea about it, first of all, we will consider the Minkowski space-time, where the rays travel in a straight line (figure \ref{fg:ms}). Whereby, we obtain four quadrants that correspond to the real image of the celestial sphere. On the other hand, when a compact object interposes between the observer and the source, the strong gravitational field that it generates deflects the path of the photons. In figure \ref{fg:gls}, we show a pictorial representation of it, where a region of the celestial sphere may appear to have been emitted from a different area due to the curved trajectories of photons. 

\subsection{Gravitational lensing in Kerr space$-$time} 
\begin{figure}
\centering
\includegraphics[scale=0.075]{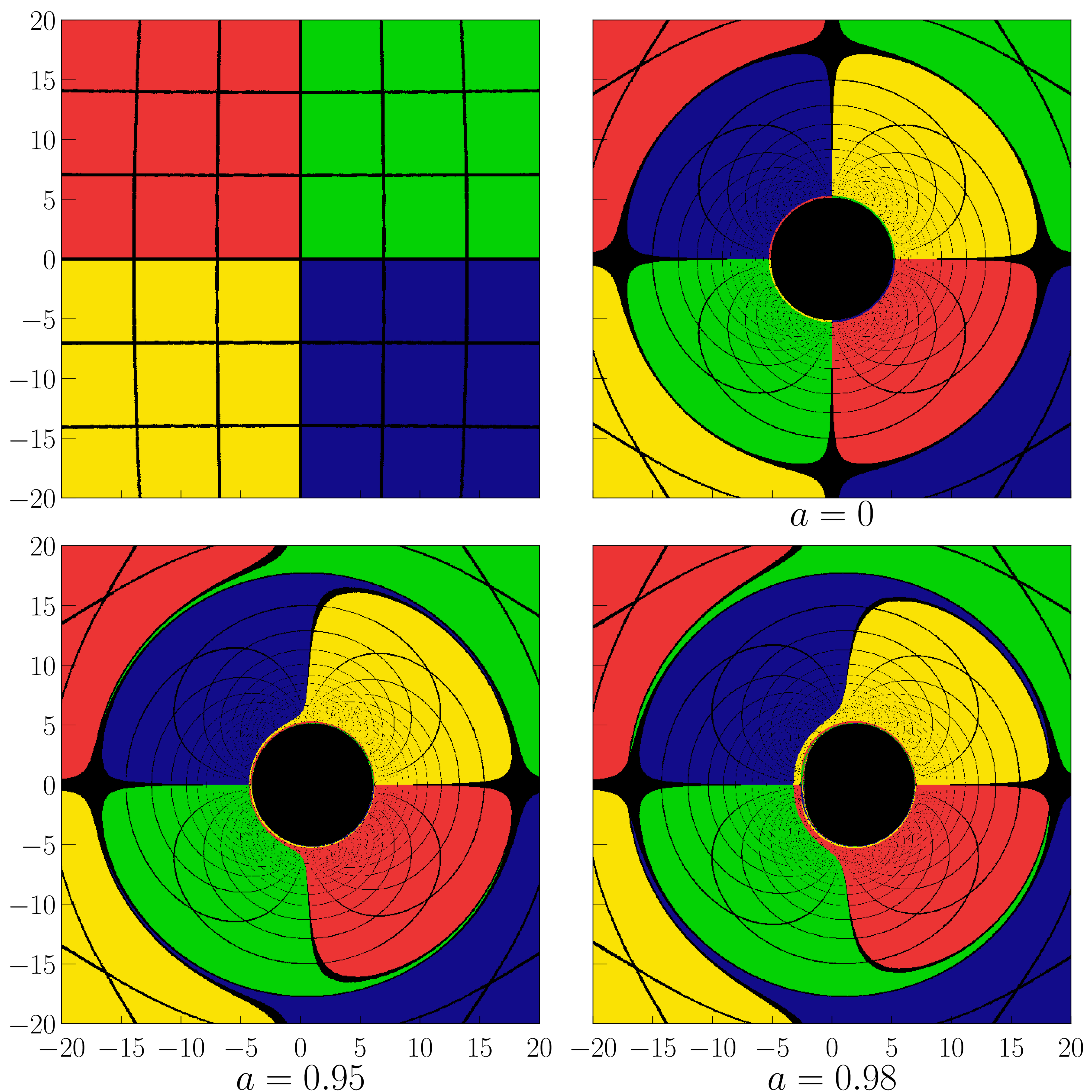}
\caption[Gravitational lensing produced by a Kerr black hole]{Gravitational lens produced by a black Kerr hole, in Boyer-Lindquist coordinates, seen from the equatorial plane at a distance $r=100$ and located at $\phi=0$. From left to right: in the top row, the Minkowski and Schwarzschild ($a=0$) space-time are shown; in the bottom row, a Kerr black hole with rotation $a=0.5$ and $a=0.98$ are observed.} \label{fg:GL}
\end{figure}

As the first application of our code, we present numerical simulations of the gravitational lens produced by a Kerr black hole for different values of the dimensionless rotation parameter $a$, seen from the equatorial plane in contrast with flat space-time. The line element that describes this geometry, in Boyer-Lindquist  coordinates $\left\lbrace t, \, r, \, \theta, \, \phi  \right\rbrace$, is given by 

\begin{gather}
\textbf{g} = - \left(1 - \frac{2Mr}{\Sigma} \right) dt \otimes dt
-
\left( \frac{4M r a \, \sin^{2}\theta}{\Sigma}\right) dt \otimes d\phi \nonumber
+
\\
\left(\frac{\Sigma}{\Delta}\right) dr \otimes dr 
+ 
\Sigma d\theta \otimes d\theta  \label{eq:Kerr} 
+ 
\\
\sin^{2}\theta \left(r^{2} + a^{2} + \frac{2Mra^{2} \sin^{2}\theta}{\Sigma} \right)d\phi \otimes d\phi, \nonumber
\end{gather}
with
\begin{equation}
\Sigma = r^{2} + a^{2} \cos ^{2}\theta \qquad \text{and} \qquad \Delta = r^{2} -2M r + a^{2},
\end{equation}
where $ a = J / M $ relates the angular momentum of rotation $ J $ and the mass $ M $ of the gravitational source, and the expression for the outer event horizon is given by 
\begin{equation}
r_H = M + \sqrt{M^{2} - a^{2}}.
\end{equation}
In the upper left row of figure \ref{fg:GL}, we show the gravitational lensing produced by the Minkowski space-time, where the black lines of the mesh are not straight due to the curvature of the sphere itself. The upper right quadrant is the non-rotating, $a=0$, corresponding to a Schwarzschild black hole. An appreciable effect of the gravitational lens is the Einstein ring, a phenomenon that appears when the observer, the compact object, and the source are aligned. Due, the light source looks like a concentric ring around the black hole. Inside this ring, the deflection angle of the photons's trajectories is big enough to generate an inversion of the colors \cite{Bohn_2015}. Additionally, a second ring can be seen near the edge of the shadow, where the images are inverted again.
The gravitational lens produced by a black Kerr hole with rotation $a=0.5$ and $a=0.98$ is shown at the bottom row of figure \ref{fg:GL}, where the Einstein ring appears as in the Schwarzschild case. However, as a result of the gravitational drag due to the black hole rotation, an asymmetry occurs in the lens. In the lower-left image, the blue and green quadrants spread along the inner edge of the ring, which generates an “ear” in yellow and red quadrants. This ear extends along with the shadow’s silhouette. These effects are more noticeable as the rotation increases. In the lower-right quadrant of figure \ref{fg:GL}, the ear spreads over the edge of the shadow and envelops it. Additionally, a consecutive succession of partial Einstein rings appears in the vicinity of the flat edge of the shadow as a consequence of extreme rotation. Furthermore, a characteristic in common is the deformation suffered by the lines of the time-space mesh, providing the notion of curvature due to the strong gravitational field.
%
\section{Thin accretion disk} \label{sec:TAD}

\begin{figure*}[htbp]
\centering
\includegraphics[scale=2]{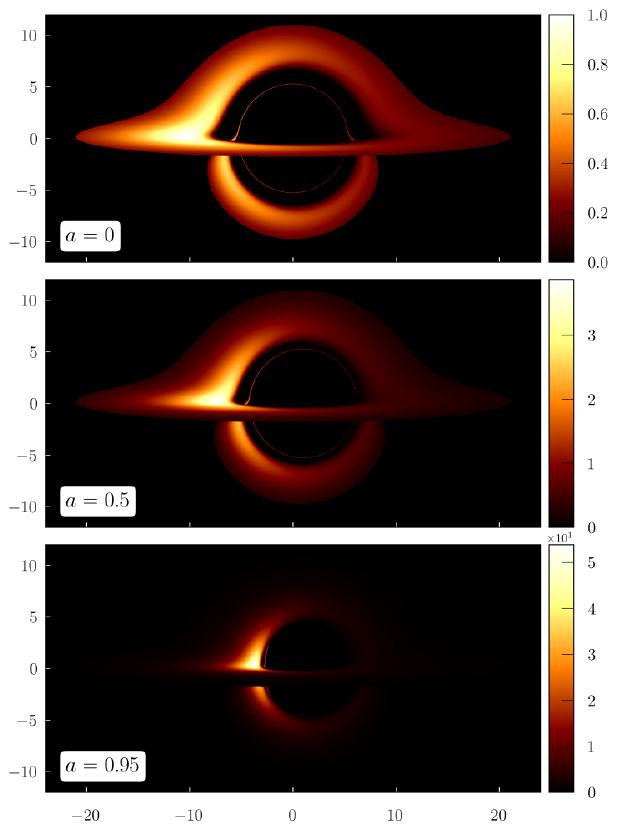}	
\caption[Thin accretion disks]{Simulation of a thin accretion disk for a black hole with different spin parameter ($a = 0, \; 0.5, \;0.998$). The observed intensity show a bright dot at the left of the disk produced by the redshift, whose location is due to the direction of rotation of the disk. It is clear that the observed intensity increases as the spin parameter increases, being an order of magnitude higher for an extreme rotation black hole ($a = 0.95$) compared to the non-rotating case ($a = 0$) and the intermediate rotation case ($a = 0.5$). The simulation parameters are: observer location at $r_0 = 1000$, $\theta_0 = 85^\circ$, $\phi_0 = 0^\circ$ and $t_0 = 0$; the limits of the disk are $r_{in} = r_{\text{isco}}$ (for every case of rotation) and $r_{out} = 20$. The specific angular momentum, $l_0$, is 2.8 for $a = 0$ and $a = 0.5$, and $l_0 = 1.8$ for $a = 0.95$. The range of the image plane is $x \in [-24,24]$ and $y \in [-12,12]$, and the resolution of the simulation is 2048$\times$1024 pixels.} \label{fg:disk}
\end{figure*}

Accretion disks are a great mechanism to determine the properties of a black hole, like its mass or spin \cite{Gurzadyan,page}. Furthermore, the emission spectrum and the radiative flux from these structures can depend on space-time generated by the compact object \citep{Takahashi},  emitting from the radio band to x-rays \citep{Krolik,Pringle}. The numerical simulations are useful to compare and interpret the observational results with different theoretical models. Next, we consider a thin accretion disk around a Kerr black hole in the equatorial plane, assuming an ideal non-self-gravitating disk moving in circular orbits ($u^r = u^\theta = 0$, with $u^\alpha$ the components of the 4-velocity for time-like particles moving on the disk). Let $\Omega$ the angular velocity and $l_0$ the specific angular momentum of time-like particles on the disk, then
\begin{equation}
\Omega := u^\phi/u^t = -\frac{g_{t\phi} + g_{\phi\phi}l_0}{g_{\phi\phi} + g_{t\phi}l_0}.
\end{equation}
 Based on our ray-tracing, the integration process for null geodesics stops when photons reach the disk, then an observed intensity is assigned to each point $(x,y)$ on the image plane. Through the Lorentz invariant $
I/\nu^3$, the observed intensity $I_{\text{obs}}$, can be expressed in terms of the emitted intensity $I_{\text{obs}}$ and the respective frequencies as follows
\begin{equation}
I_{\text{obs}} = g^3 I_{\text{em}},
\end{equation}
with
\begin{equation}
g = \nu_{\text{obs}}/\nu_{\text{em}} = (1+z)^{-1} = \frac{\mathcal{P}_\beta {\mathcal{U}}^\beta}{p_\alpha u^\alpha}, \label{eq:redshift}
\end{equation}
the redshift factor due to the gravitational spectral shift and the Doppler effect (recalling that calligraph letters means physical magnitudes measured by the observer). 
Now, employing the normalization condition
\begin{equation}
g_{\alpha\beta}u^\alpha u^\beta = -1, \label{eq:norm}
\end{equation}
the equation \ref{eq:redshift} can be written as
\begin{equation}
g = p_t\left(1+\Omega\frac{p_\phi}{p_t}\right)\left(-g_{tt}-g_{\phi\phi}\Omega^2 - 2\Omega g_{t\phi}\right)^{-1/2}.
\end{equation}
As result of our numerical simulation, in the figure \ref{fg:disk}, we show the image of a Kerr black hole surrounded by a thin accretion disk for different dimensionless spin parameter ($a=0, \; 0.5, \; 0.95$) with a map for the observed intensity using the disk model proposed in \citep{page}, where it is evident that high rotation increases the intensity measured by a distant observer. We can appreciate a bright spot located at the left of the disk as an effect of the redshift, which decreases its size as rotation increases, concentrating the maximum observed intensity in a small region. On the other hand, the bend of light due to the extreme gravitational field allows us to see the back of the disk, which should be hidden by the black hole (a deeper discussion and more details are carried out by Luminet \citep{luminet}). Additionally, when considering the disk as a bright source, it is possible to appreciate how the edge of the shadow is delimited by a thin layer of light coming from the disk. 

\lipsum[500]
\lipsum[354]

\section{Conlusions} \label{sec:CON}

In this paper, we present \texttt{OSIRIS}, new stable ray-tracing FORTRAN code capable of efficiently compute null-geodesics in stationary and axisymmetric space$-$times. \texttt{OSIRIS} incorporates general expressions for the impact parameters relating to the measurements of an observer located in the vicinity of the gravitational source and another at infinity. As a first application, we simulate the image of a rotating black hole seen by a distant observer and study (qualitatively) the effect of the gravitational dragging in the shadow and gravitational lensing.

The image of a compact object is obtained by solving the motion equations for photons in each point of the observer's screen. We implement four different stable integration schemes with adaptive step: Runge-Kutta Fehlberg 45, Runge-Kutta Cash-Karp 45, Runge-Kutta Dormand-Prince 45, and a Bulirsch-Stoer algorithm. We analyze the error in the Hamiltonian constraint for null particles ($H=0$) and find that the RK Dormand-Prince preserves the constraint better than a part in $10^{-11}$, while the other schemes produce considerably larger errors. Although this method requires more computational time than the RK Fehlberg and RK Cash-Karp, especially when dealing with high-resolution simulations, the time differences between the schemes are negligible for reasonable resolutions. Therefore, the RK Dormand-Prince emerges as the best option for the main goal of \texttt{OSIRIS} (photons dynamics around compact objects). Additionally, it should be noted that even though OSIRIS is parallelized neither with MPI nor with CUDA, the computation times are similar to those obtained by other codes programmed with these parallel computing platforms. 

\texttt{OSIRIS} can simulate accretion disks around black holes, from which it is clear that the apparent form of the black hole and the observed intensity depend on the space-time that is considered. For this reason, develop theoretical models for accretion disk surrounding black holes is of great importance when comparing observational results with the computational simulations, since from these it is possible to obtain information about the space-time itself.

\section*{Acknowledgments}
F. D. L-C acknowledge support from Vicerrectoría de Investigación y Extensión - Universidad Industrial de Santander, Colombia under Grant No. 2493., J. M. Velásquez-Cadavid and J.A. Arrieta-Villamizar acknowledge to Universidad Industrial de Santanter for financieral support during the construction of this article. 
%

%
\appendix
\section{Time-Like Geodesics}\label{appendix:A}

\begin{figure*}[htbp]
\centering
\includegraphics[scale=1]{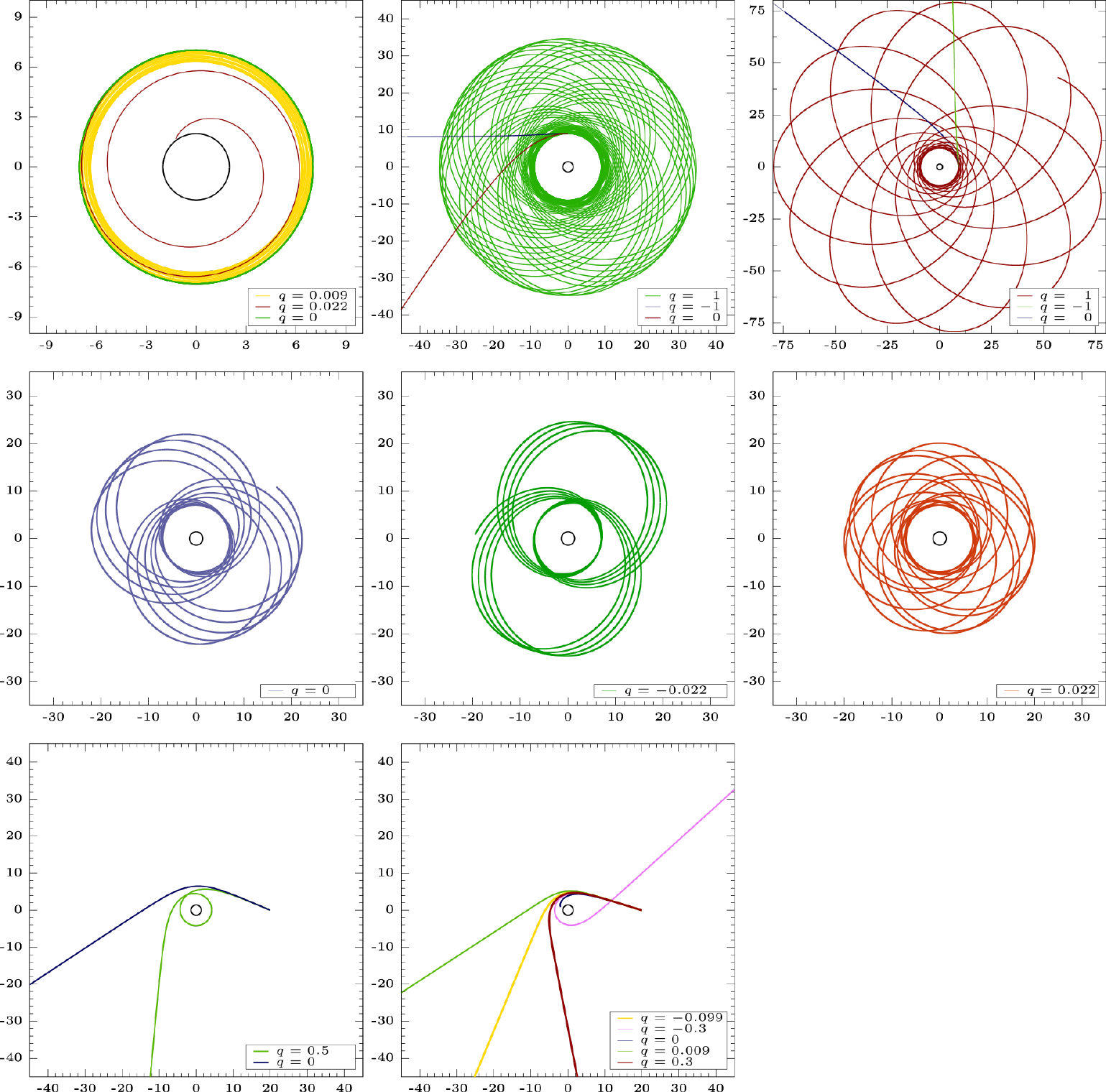} \caption{Time-like geodesics for different values of the quadrupole in the equatorial plane ($\theta = \pi/2$). In the first row, there are unbounded Schwarzschild orbits, with $\dot{r} = 0$. In the second row, there are bounded Schwarzschild orbits with $\dot{r} = 0$. Finally, in the third row, the particles with $\dot{r} \neq 0$ scape toward infinity.} \label{fg:timelike}
\end{figure*}

\texttt{OSIRIS} also was tested solving geodesics equations for test particles around a mass source with deformation. For this purpose, we chose the simplest static and axially symmetric solution to the Einstein equations with a non-vanishing quadrupole moment, which corresponds to the $q$-metric \citep{HQuevedo1}. In spherical coordinates, the metric tensor for this space-time is 
\begin{align}\label{eq:q-metric}
  \centering
  \mathbf{g} &= \left(1 - \frac{2m}{r}\right)^{1 + q}\operatorname{d}{t} \otimes \operatorname{d}{t} - \left(1 - \frac{2m}{r}\right)^{ - q} \times \nonumber \\ \nonumber
            &  \Bigg[ \Bigg( \operatorname{d}{r} \otimes \operatorname{d}{r}\Bigg(1 - \frac{2m}{r}\Bigg)^{-1}
       + r^2 \operatorname{d}{\theta} \otimes \operatorname{d}{\theta} \Bigg) \nonumber \\ & \times \left( 1 + \frac{m^2\sin^2{\theta}}{r^2 - 2mr}  \right)^{-q(2 +q)} + r^2 \sin^2 {\theta} \operatorname{d}{\varphi} \otimes \operatorname{d}{\varphi} \Bigg],
   \end{align}   
and describes a deformed mass distribution, where the parameter $q$ determines the deformation of the source. For values of $q$ in the interval $(-1, -1+\sqrt{3/2})\setminus\{0\}$, the source corresponds to naked singularities without event horizons \citep{HQuevedo2}. Besides, for $q= 0$, this metric reduces to Schwarzschild space-time. In figure \ref{fg:timelike}, we show time-like geodesics in the equatorial plane for different values of $q$. In the first row, we appreciate how the influence of $q$ transforms unbounded Schwarzschild trajectories into bounded ones for particles with non-initial radial velocity ($\dot{r} = 0$). In the second row, we show a Schwarzschild bounded orbit and the geodesics with $\dot{r} = 0$ around a prolate ($q<0$) and oblate ($q > 0$) source. We can see that the quadrupole only affects the structure of the trajectory. Finally, in the third row, we present particles with $\dot{r} \neq 0$. In this case, all of them scape toward infinity, and the parameter $q$ only defines the scape direction.
\lipsum[500]
\lipsum[354]

\end{document}